\documentclass[12pt,amsfonts]{article}

\usepackage{amssymb}
\usepackage{bm}
\usepackage{graphics}

\newcommand{\be}{\begin{equation}}
\newcommand{\ee}{\end{equation}}
\newcommand{\ba}{\begin{array}}
\newcommand{\ea}{\end{array}}
\newcommand{\bqa}{\begin{eqnarray}}
\newcommand{\eqa}{\end{eqnarray}}
\newcommand{\bea}{\begin{eqnarray}}
\newcommand{\eea}{\end{eqnarray}}

\begin{document}

\newtheorem{defi}{Definition}[section]
\newtheorem{lem}[defi]{Lemma}
\newtheorem{prop}[defi]{Proposition}
\newtheorem{theo}[defi]{Theorem}
\newtheorem{rem}[defi]{Remark}
\newtheorem{cor}[defi]{Corollary}

\newcommand{\qed}{\hfill $\Box$\vspace{.5cm}\medskip}


\title{Inhomogeneous Heisenberg Spin Chain and Quantum Vortex Filament as Non-Holonomically Deformed NLS Systems}

\author{Kumar Abhinav$^{\dag}$\footnote{kumar.abhinav@bilkent.edu.tr} 
and Partha Guha$^{\ddag}$\footnote{partha@bose.res.in}\\
\\
$^\dag$\,Dept. of Physics, Bilkent University,\\ 06800 \c{C}ankaya, Ankara, Turkey\\
\\
$^\ddag$\,S. N. Bose National Centre for Basic Sciences,\\ JD Block, Sector III, Salt Lake, Kolkata 700106, India}


\date{}
\maketitle

\abstract{Through the Hasimoto map, various dynamical systems can be mapped to different integrodifferential generalizations
of Nonlinear Schr\"odinger (NLS) family of equations some of which are known to be integrable. Two such
continuum limits, corresponding to the inhomogeneous XXX Heisenberg spin chain
[Balakrishnan, J. Phys. C {\bf 15}, L1305 (1982)] and that of a thin vortex filament moving in a
superfluid with drag [Shivamoggi, Eur. Phys. J. B {\bf 86}, 275 (2013) 86; Van Gorder, Phys. Rev. E
{\bf 91}, 053201 (2015)], are shown to be {\it particular} non-holonomic
deformations (NHDs) of the standard NLS system involving generalized parameterizations. Crucially, such
NHDs of the NLS system are restricted to specific spectral orders that exactly complements NHDs
of the original physical systems. The specific non-holonomic constraints associated with
these integrodifferential generalizations additionally posses distinct semi-classical signature.}

\bigskip


\bigskip

{\bf Keywords and Keyphrases}: Heisenberg Spin Chain, Non-Holonomic deformation, Non-linear Schr\"odinger equation, Integro-partial differential equations.

\bigskip


\section{Introduction}
The dynamics of three-dimensional continuous systems can adequately be realized
in the Frenet-Serret coordinate system having a normalized basis. The latter has the geometric interpretation
of a generic three-dimensional moving curve with torsion and curvature. This correlates the
original dynamical parameters to spatial geometry, providing an elegant formulation. The Hasimoto map
\cite{N4}, 

\be
q(s,u)=\kappa(s,u)\exp\left\{i\int_{-\infty}^s\tau(s',u)ds'\right\},\label{N5}
\ee  
further relates these geometric parameters to a complex amplitude function $q(s,u)$ leading to
second-order differential equations of the NLS type. Here $s$ and $u$ are respective space and
time parameters of the Frenet-Serret space having curvature $\kappa$ and torsion $\tau$.\\

In addition to various continuous ones some discrete physical systems can also posses
such continuum descriptions through the Frenet-Serret representation subjected to suitable approximations such as
long wavelength and perturbative limits. We consider two such systems of very distinct physical origins
and identify them as non-holonomic deformations (NHD) \cite{NHD1,KunduNHD} of the standard NLS system
across the Hasimoto map. A NHD is obtained specifically by deforming an integrable system {\it without}
hampering its scattering profile, thereby necessarily imposing certain additional constraints on the
extended system \cite{KunduNHD}. If the final system remains integrable the deformation is sub-classified
as semiholonomic \cite{NHD1}, which is the case for one of the systems under consideration. It is found
that a definite class of such deformed NLS equations corresponds to the particular systems
under consideration.

\paragraph*{}As the first system we consider the Heisenberg system of $N$ interacting spin-$1/2$
fermions in one dimension described in the $2^N$ dimensional product space ${\cal N}=\bigotimes_{j=1}^{N}h_j$,
where $h_j$s are two-dimensional vector spaces over ${\Bbb C}$. The
standard basis for $h_j$s are $e_{j}^{+} = (1~0)^T$ and $e_{j}^{-} = (0~1)^T$, spanned by the Pauli
matrices $\sigma_{x,y,z}$. The corresponding Heisenberg Hamiltonian is given by \cite{01},

$$ H = -\frac{1}{2}\sum_{j=1}^{N}\left(J_x\sigma_{j}^{x}\sigma_{j+1}^{x} + J_y\sigma_{j}^{y}\sigma_{j+1}^{y}
+ J_z\sigma_{j}^{z}\sigma_{j+1}^{z}\right), $$
representing nearest neighbor interaction between on-site spins ${\cal S}_i=\hat{k}\sigma_i^k$
with $J_{x,y,z}$ being real
constants. For $J_x = J_y = J_z = J$ this Hamiltonian reduces to the XXX Hamiltonian
$H=-(J/2)\sum_{i=1}^N{\cal S}_i\cdot{\cal S}_{i+1}$ up to a constant, with isotropic interaction strength
$J$. Here, ${\cal S}_{i}\cdot{\cal S}_{i+1} = {\boldsymbol\sigma}_i\cdot{\boldsymbol\sigma}_{i+1}$.
The sign of $J$ represents the corresponding magnetic orientation. The anisotropy in the internal
$SU(2)$ spin sub-space leads to XXZ or XYZ models. Under the continuum limit limit of vanishing
lattice constant $a\to0$ and normalization $Ja^2=1$ \cite{N1}, The equation of motion (EOM)
${\cal S}_{i,u}=J{\cal S}_i\times\sum_{j=i-1}^{i+1}{\cal S}_j~(\hbar=1)$ for the XXX system reduces
to a semi-classical one \cite{N2}:

\be
{\mathbf t}_u={\mathbf t}\times{\mathbf t}_{ss},\label{N3}
\ee
that parameterizes a moving curve in ${\Bbb R}^3$. The Frenet-Serret representation is invoked through
the tangent unit vector ${\mathbf t}(u,s)$ that doubles as the continuum limit of ${\cal S}_i$.
Subsequently, a Hasimoto map (Eq. \ref{N5}) leads to the standard NLS system \cite{N2}:

\be
q_u-iq_{ss}-2i\eta\vert q\vert^2q=0,\label{N4}
\ee
having energy and momentum densities $\kappa^2$ and $\kappa^2\tau$.\\

In fact, Hasimoto originally obtained the NLS system as a continuum limit of a moving
vortex filament described by Eq. \ref{N3} \cite{N4}, the latter corresponding to the XXX HSC, with a
one-to-one correspondence between respective soliton solutions \cite{N2}. As a generalization, the
inhomogeneous XXX model ($J\to J_i$) was similarly mapped to an integrodifferential
modification of the NLS system \cite{R01,N3} which was shown to be integrable \cite{N5} with a geometric
interpretation \cite{2}. Therefore, the proposed NHD relating the NLS system to this integrodifferential
generalization is a semiholonomic one \cite{NHD1}.

\paragraph*{}The second system under consideration corresponds to a thin vortex filament moving in
a binary mixture of superfluid $^4{\rm He}$ and a classical fluid with a velocity ${\bf v}$.
In 1956, Hall and Vinen \cite{HV1,HV2} developed a coarse-grained hydrodynamic
equation capable of describing a superfluid having a continuously distributed vorticity.
Their equations are valid only if the typical length scale of the problem is much
larger than the inter-vortex spacing. Later, Bekarevich and Khalatnikov \cite{BK} presented a more
elaborated version of these equations which is now known as the Hall-Vinen-Bekarevich-Khalatnikov
(HVBK) equation:
$$
{\boldsymbol v}=\kappa{\boldsymbol t}\times{\boldsymbol n}+\alpha{\boldsymbol t}\times\left({\boldsymbol U}-\kappa{\boldsymbol t}\times{\boldsymbol n}\right)-\alpha'{\boldsymbol t}\times\left[{\boldsymbol t}\times\left({\boldsymbol U}-\kappa{\boldsymbol t}\times{\boldsymbol n}\right)\right],
$$
with dimensionless normal velocity ${\bf v}$, local curvature $\kappa$ and dimensionless small friction
coefficients $\alpha,~\alpha^{\prime}$. The Hasimoto type formulation of this problem was given by
Shivamoggi \cite{Shiva}, coined as quantum Hasimoto map.\\

Thus the moving curve described by Eq. \ref{N3} can be generalized to a quantized thin
vortex filament in a superfluid medium \cite{NN1}, given by an equation under the local induction
approximation (LIA) which is accompanied by friction at finite temperature, rendering it non-integrable.
This, along with the inhomogeneous XXX system, represents two integrodifferential NLS generalizations obtained
from motion of generic Frenet-Serret curves through the Hasimoto map. In this work, we show that
both these extended NLS systems can be realized as particular restricted non-holonomic deformations of the standard
NLS system with generic parameterizations. In fact, this is the key feature of the paper which
has been overlooked till now. The XXX model corresponds to a NHD confined to a very specific
{\it spectral region} containing all possible contributions to modification of the dynamics. Further,
there is always an additional amplitude-phase correlation represented by the constraint itself, as a
strong semi-Classical signature.

\paragraph*{}In the following, the NHDs of the NLS system leading to those describing inhomogeneous XXX
model and quantum vortex filament in superfluid at finite temperature are discussed in Section 2.
Section 3 depicts the spectral restriction for the validity of such deformations, followed by the
observation of the semi-Classical nature therein. We conclude in section 4, emphasizing on the possible
introduction of $SU(2)$ anisotropy, along with possible extension of the present procedure to a larger
class of systems.

\section{Extended dynamical Systems as Specific NHDs of NLS Systems}\label{SXXX}
The NHD of an integrable system is achieved through perturbation by virtue of {\it additional
constraints} \cite{NHD1}. The constraints can modify existing terms in or introduce new ones in
the dynamical equation. Mathematically, such deformations are introduced by exclusively modifying the
temporal component $B(\lambda)$ of the Lax pair in order to keep the scattering data unchanged
\cite{KunduNHD}, although the temporal evolution gets modified. In some cases integrability is
preserved which are known as semiholonomic deformations \cite{NHD1}. They correspond to a certain
form of self-consistent source equation pertaining to the original integrability property. Retainment
of integrability requires the non-holonomic constraints to be affine in the velocities so that the
deformed dynamical equation does not have explicit velocity dependence and thereby integrable. Deforming
the temporal Lax component serves this purpose as in absence of its time-derivative in the flatness
condition, particular cases can retain velocity independent dynamics in accordance with Ref. \cite{NHD1}.
Of course in other cases of NHD with a generic
modification to $B(\lambda)$, non-integrability results starting from an integrable system. We
will consider one example of each of these two cases with the same undeformed integrable origin, the
NLS system. In general, higher derivative
hierarchies arise as a natural outcome of these deformations, either through recursive higher order
constraints while keeping the order of the perturbed system same, or by fixing the constraints in
the lowest order and thereby increasing the differential order of the original equation itself. If
they correspond to a semiholonomic structure, these hierarchies are integrable too.\\

In this section, we demonstrate that the integrodifferential continuum representations of two
different dynamical systems, obtained through Hasimoto map, can be viewed as specific NHDs of a
{\it generalized} NLS family. Namely, we consider the NLS-like continuum limits of inhomogeneous
Heisenberg spin chain \cite{N3,N5} and the local induction approximation (LIA) of a moving thin
vortex filament \cite{NN1}, with the prior retaining integrability over the deformation. It is found
that such systems are characterized by deformations restricted to particular spectral domains, with
additional semi-Classical characteristics ({\it i. e.}, quantum signature of the discrete analogue).

\subsection{Inhomogeneous XXX Model as a NHD of NLS System}\label{2.1}
To discuss the NHD of NLS system we consider the following representation of the generalized
NLS Lax pair \cite{FL}:

\bea
&&A=-i\lambda\sigma_3+\rho^*q^*\sigma_++\rho q\sigma_- \quad{\rm and}\nonumber\\
&&B=i\left(2\lambda^2-\eta\vert q\vert^2\right)\sigma_3-\left(2\lambda\rho^*q^*+i\rho^*q^*_x\right)\sigma_+\nonumber\\
&&\qquad-\left(2\lambda\rho q-i\rho q_x\right)\sigma_-,\label{13}
\eea
in the $sl(2)$ representation built on $SU(2)$ algebra: $[\sigma_3,\sigma_\pm]=\pm2\sigma_\pm$,
$[\sigma_+,\sigma_-]=\sigma_3$. The usual NLS system (Eq. \ref{N4}) is obtained from the zero curvature
condition (ZCC):

$$F_{tx}=A_t-B_x+[A,B]=0.$$
Here $\eta=-\vert\rho\vert^2$ for consistency following ${\cal O}\left(\lambda^0\right)$ contribution
to the ZCC. The ${\cal O}\left(\lambda\right)$ sector of the ZCC necessitates $\rho$ (thus $\eta$ also)
to be a constant. Therefore local coefficients are not allowed by the integrability structure itself
and thereby the prohibits inhomogeneity. \\

In order to invoke a more generalized NLS system with {\it local} coefficients that can represent
the {\it inhomogeneous} XXX model in the continuous limit, it is therefore natural to adopt a NHD.
This makes sense as NHD can effect modified time-evolution that allows for
compensating space evolution maintaining integrability and preserving the scattering data. For this
purpose we consider the
discrete inhomogeneous XXX Heisenberg spin chain:

\be
H'=-\sum_i\rho_i{\cal S}_i\cdot{\cal S}_{i+1},\label{N6}
\ee 
with {\it on-site} ferromagnetic parameter $\rho_i$. It has already been shown \cite{N3} that the continuum
limit of this system is an integrodifferential generalization of the NLS system:

\be
q_t-i\left(\rho q\right)_{xx}-2i\rho q\vert q\vert^2-2iq\int^x_{-\infty}\rho_{x'}\vert q\vert^2dx'=0.\label{N7}
\ee
The integrability of this system was established through an extended inverse-scattering analysis to
incorporate $x$-dependence of the coupling coefficient $\rho (x)$ by Balakrishnan \cite{N5}. Therein,
explicit soliton solutions are obtained for a wide class of functions as $\rho (x)$ referring to inhomogeneous
physical interactions that further included generalized Gelfand-Levitan equation.
Except for the last integrodifferential term, the above equation resembles with a
{\it focusing} type of NLS system, represented by Eq. \ref{N4}\footnote{The defocusing case
corresponds to substituting $q^*=-q^*$ in the Lax pair of Eq. \ref{13}, changing the sign of the
nonlinear term in the NLS equation. Though both focusing and defocusing NLS systems are integrable with
distinct solution spaces, only the focusing-type inhomogeneous `extension' corresponds to the
inhomogeneous XXX model and is known to be integrable \cite{N5}.}. Therefore it is reasonable to
expect that the above integrable equation is a particular NHD of the usual NLS system. It makes additional
sense since the target equation has local parameters not allowed in the standard NLS Lax pair as per
the ZCC, a condition most likely to be relaxed by introducing deformation parameters.

\paragraph*{}We perform NHD of the standard NLS system through introducing the following perturbation
in the temporal Lax component as:

\bea
&&B\to B'= B+\delta B,\nonumber\\
&&\delta B=\frac{i}{2}\left[f_3\sigma_3+f_+\sigma_++f_-\sigma_-\right],\label{N011}
\eea
with local, time-dependent parameters $f_{\pm,3}$ accompanied by generalizations $\eta=\eta(x,t)$
and $\rho=\rho(x,t)$. Then the ZCC, with temporal Lax component $B'$, leads to independent
equations corresponding to each linearly independent $SU(2)$ generator. The ${\cal O}\left(\lambda^0\right)$
contributions are:

\bea 
&&\sigma_+:\quad \rho^*\left(q^*_t+iq^*_{xx}+2i\eta\vert q\vert^2q^*\right)+i\rho_x^*q^*_x+\rho^*_tq^*\nonumber\\
&&\qquad\quad-\frac{i}{2}\left(f_{+,x}+2\rho^*q^*f_3\right)=0,\nonumber\\
&&\sigma_-:\quad \rho\left(q_t-iq_{xx}-2i\eta\vert q\vert^2q\right)-i\rho_xq_x+\rho_tq\nonumber\\
&&\qquad\quad-\frac{i}{2}\left(f_{-,x}-2\rho qf_3\right)=0,\nonumber\\
&&\sigma_3:\quad \left(\eta\vert q\vert^2\right)_x+\vert\rho\vert^2\vert q\vert^2_x\nonumber\\
&&\qquad\quad-\frac{1}{2}\left(f_{3,x}+\rho qf_+-\rho^*q^*f_-\right)=0,\label{N012}
\eea
The spectral order of the perturbation, which is also $\lambda^0$ in this case,
is of crucial importance which will be elaborated later. At ${\cal O}\left(\lambda^1\right)$ of the
ZCC we get:

\bea
&&\sigma_+:\quad \rho_x^*q^*+\frac{1}{2}f_+=0 \quad{\rm and}\nonumber\\
&&\sigma_-:\quad \rho_xq-\frac{1}{2}f_-=0.\label{N013}
\eea
It is easy to see that without $f_\pm$ (standard NLS system), the parameter $\rho$ becomes
space-independent (usual NLS case). As $f_\pm$ get fixed by the above equations, from the last of
Eq.s \ref{N012}:

\be
f_3=2\left(\eta+\vert\rho\vert^2\right)\vert q\vert^2+T(t),\label{N014}
\ee
where $T(t)$ is the space-integration constant that has pure time-dependence. Therefore from Eq.s
\ref{N013} and \ref{N014} all three deformation parameters get fixed, finally leading to the deformed
system,

\be
ip_t+p_{xx}-2p\vert p\vert^2=2T(t)p,\quad p=\rho q.\label{R01}
\ee
This essentially is the defocusing NLS system. The time-dependent source term, which is a
trivial integration constant, can be set equal to zero. It is evident that although $\rho$ can still
be time-dependent for $f_\pm=0$, the overall scaling $q\to p=\rho q$ does not allow any time-dependent
interaction, thereby preserving integrability. However, the above system is not the desired one
(Eq. \ref{N7}) and there is no additional higher-order constraint
equation signifying NHD. Essentially an NHD only of ${\cal O}\left(\lambda^0\right)$ will
always have deformation parameters $f_{3,\pm}$ completely determined in terms of undeformed parameters
($q$, $\rho$ and $\eta$).This leaves no room for additional constraint dynamics and only a system
similar to the undeformed one is obtained. The dynamical variable merely attains local scaling while the
nonlinear coupling parameter gets scaled to unity.\\ 

To obtain the desired equation one needs to go beyond ${\cal O}\left(\lambda^0\right)$
in NHD. The NLS system is obtained from the Lax pair of Eq.s \ref{13} at
${\cal O}\left(\lambda^0\right)$ with additional conditions coming from ${\cal O}\left(\lambda^1\right)$.
Thus the only substantial contribution can come from {\it additional} NHDs up to
${\cal O}\left(\lambda^{-1,1}\right)$. Among them a ${\cal O}\left(\lambda^1\right)$ deformation
of the form $\delta B= \lambda g_i\sigma_i,\quad i=3,\pm$ would lead only to,

\be
ip_t+p_{xx}+2p\vert p\vert^2=2T(t)p-\frac{i}{2}G(t)p_x,\label{RR1}
\ee
with time-dependent source term, again with no constraints. This particular deformation would contribute at
${\cal O}\left(\lambda^{1,2}\right)$, with the ${\cal O}\left(\lambda^2\right)$ sector devoid of original
parameters ($q,\rho,\eta$), thereby severely restricting this particular deformation. The above system is
different from the previous one, with the second
source term $G(t)$ being the ${\cal O}\left(\lambda^1\right)$ deformation parameter which is free.
However still no integrodifferential modification possible with such a temporal deformation.

\paragraph*{}The desired result is finally obtained for an ${\cal O}\left(\lambda^{0,-1}\right)$
deformation:

\be
\delta B=\frac{i}{2}\left(f_i+\lambda^{-1}h_i\right)\sigma_i,\quad i=3,\pm.\label{R02}
\ee
Unlike the ${\cal O}\left(\lambda\right)$ case, now the ${\cal O}\left(\lambda^{-1}\right)$ contribution
directly modifies the EOM as,

\bea 
&&\sigma_+:\quad \rho^*\left(q^*_t+iq^*_{xx}+2i\eta\vert q\vert^2q^*\right)+i\rho_x^*q^*_x+\rho^*_tq^*\nonumber\\
&&\qquad\quad-\frac{i}{2}\left(f_{+,x}+2\rho^*q^*f_3\right)=-h_+,\nonumber\\
&&\sigma_-:\quad \rho\left(q_t-iq_{xx}-2i\eta\vert q\vert^2q\right)-i\rho_xq_x+\rho_tq\nonumber\\
&&\qquad\quad-\frac{i}{2}\left(f_{-,x}-2\rho qf_3\right)=h_-,\nonumber\\
&&\sigma_3:\quad \left(\eta\vert q\vert^2\right)_x+\vert\rho\vert^2\vert q\vert^2_x\nonumber\\
&&\qquad\quad-\frac{1}{2}\left(f_{3,x}+\rho qf_+-\rho^*q^*f_-\right)=0,\label{R03}
\eea
at ${\cal O}\left(\lambda^0\right)$. The ${\cal O}\left(\lambda^1\right)$ sector remains same as
Eq.s \ref{N013}. The new contributions appear at ${\cal O}\left(\lambda^{-1}\right)$ as,

\bea
&&h_{3,x}=\rho^*q^*h_--\rho qh_+,\quad h_{+,x}=-2\rho^*q^*h_3\nonumber\\
{\rm and}\quad &&h_{-,x}=2\rho qh_3,\label{R04}
\eea
Thus $h_i$ are mutually constrained and $h_3$ appears only in these constraint equations. Moreover,
the EOMs impose that $h_-=-h_+^*$. Thus the inhomogeneous source terms $h_\pm$ can be fixed
through defining $h_3$ judiciously. Assuming this very case, $h_\pm$ can suitably be chosen as:

\bea
&&h_-=\left\{(\rho-1)q\right\}_t+2i(1+\vert\rho\vert^2)\rho q\vert q\vert^2\nonumber\\
&&\qquad+i\rho qT(t)+2iq\int^x_{-\infty}\rho_{x'}\vert q\vert^2dx'\equiv-h^*_+.\label{R05}
\eea
to yield the desired result of Eq. \ref{N7} as,

\be
q_t-i\left(\rho q\right)_{xx}-2i\rho q\vert q\vert^2-2iq\int^x_{-\infty}\rho_{x'}\vert q\vert^2dx'=0.
\ee
The fact that Eq. \ref{N7} {\it is} integrable \cite{N5} instantly testify the present deformation
to be semiholonomic \cite{NHD1}. This is further indicated by the fact that although trivial, the
time-dependent factor $T(t)$ is removed from the EOM through the choice in Eq. \ref{R05}. As a 
formal validation, the third deformation parameter can now be obtained as:

\bea
&&2\rho qh_3=\{(\rho-1)q\}_{xt}+2i\left\{\left(1+\vert\rho\vert^2\right)\rho q\vert q\vert^2\right\}_x+i(\rho q)_xT(t)\nonumber\\
&&\qquad\qquad+2iq\rho_x\vert q\vert^2+2iq_x\int^x_{-\infty}\rho_{x'}\vert q\vert^2dx'.\label{R06}
\eea
The corresponding constraint is inferred by re-combining Eq.s \ref{R04} as,

\be
h_{3,xx}=4\vert p\vert^2h_3+p^*_xh_--p_xh_+,\quad p=\rho q,\label{R07}
\ee 
which is clearly {\it fourth} order in derivatives. This is characteristic of NHD with the constraint
being of higher order in derivatives than the EOM, restricting only the solution-space but
{\it not} the dynamics.

\paragraph*{}The ${\cal O}\left(\lambda^0\right)$ deformation ($f_{3,\pm}$) is necessary for
maintaining Eq.s \ref{N013} in order to keep the original parameter $\rho$ local, which cannot be obtained
otherwise. The last deformation is the only possible NHD leading to the desired result and for that
fact to have any non-trivial local modification of the NLS system. This aspect will be discussed in detail
in the next section.

\subsection{Quantum Vortex Filament with Friction as a NHD of NLS System}
Now we consider an example of a non-integrable integrodifferential equation realized as generic NHD
of the NLS system, which also corresponds to a physical system. On considering
the frictional force exerted by the normal fluid (or quasi-particles) on a scattering vortex
line, its self-advection velocity is given by the HVBK equation \cite{NN2}:

\bea
&&{\boldsymbol v}=\kappa{\boldsymbol t}\times{\boldsymbol n}+\alpha{\boldsymbol t}\times\left({\boldsymbol U}-\kappa{\boldsymbol t}\times{\boldsymbol n}\right)\nonumber\\
&&\qquad-\alpha'{\boldsymbol t}\times\left[{\boldsymbol t}\times\left({\boldsymbol U}-\kappa{\boldsymbol t}\times{\boldsymbol n}\right)\right],\label{NN03}
\eea
according to the LIA. In the adopted Frenet-Serret basis ${\boldsymbol U}$ is the normal fluid velocity,
${\boldsymbol v}$ is the filament velocity and ${\boldsymbol n}$ is the unit normal to the filament
with tangent ${\boldsymbol t}$. Here, $(\alpha,\alpha')$ are dimensionless friction coefficients. This equation,
through a Hasimoto map (Eq. \ref{N5}), is mapped to the extended NLS-type system \cite{NN1}:

\bea
&&q_t=iA(t)q+\left\{i(1-\alpha')+\alpha\right\}q_{xx}+\left\{\frac{i}{2}(1-\alpha')-\alpha\right\}q\vert q\vert^2\nonumber\\
&&\qquad-\frac{\alpha}{2}q\int_0^x\left(qq^*_{x'}-q^*q_{x'}\right)dx',\label{NN04}
\eea
with drag coefficient $A(t)$. This equation possesses Stokes wave solution with expected decay. The present system
is naturally non-integrable owing to the explicit time dependence and thus found to
be a generic NHD (not semiholonomic) of the NLS system in the following.

\paragraph*{}In order to realize the above system as a NHD of NLS systems, we can take queues from the
inhomogeneous HSC. Since there is no local parameter other than the dynamical variable $q$ in the present
case, the ${\cal O}\left(\lambda^0\right)$ part of the NHD is reduced only to $(i/2)f_3\sigma_3$ as
$f_\pm=0$ following Eq. \ref{N013}. The dynamical equation (the second of Eq.s \ref{R03}) then takes
the simpler form,

\be
q_t-iq_xx+2i\vert\rho\vert^2q\vert q\vert^2+iT(t)q=\frac{h_-}{\rho},\label{RR03}
\ee
where $\rho$ is a complex number. Then subjected to the choice,

\bea
&&h_-=\rho\left[2i\vert\rho\vert^2+\frac{i}{2}(1-\alpha')-\alpha\right]q\vert q\vert^2\nonumber\\
&&\qquad-\frac{\alpha}{2}\rho q\int_0^x\left(qq^*_y-q^*q_y\right)dy\equiv-h_+^*,\label{RR04}
\eea
aided by the coordinate re-scaling,

\be
x\to\left(i-i\alpha'+\alpha\right)^{-1/2}x,\label{RR05}
\ee
the identification $T(t)=-A(t)$ yields the desired result (Eq. \ref{NN04}). As seen in the case of inhomogeneous
HSC, NHD of ${\cal O}\left(\lambda^{0,-1}\right)$ suffices for the present system also, being derived
from the same NLS system. This spectral bound property will be elaborated in the next section.

\paragraph*{}It is clear that if one starts with local coupling parameters in the parent NLS system, the
corresponding NHD variables will become more extensive. A straight-forward derivation in this line,
with $f_\pm\neq 0$, leads to the dynamical equation,

\be
p_t-ip_{xx}+2ip\vert p\vert^2+ipT(t)=h_-,\quad p=\rho q\label{RR06}
\ee
To obtain Eq. \ref{NN04}, the required choice of parameter is,

\bea
&&h_-=(p-q)_t-\left[p-\{\alpha+i(1-\alpha')\}q\right]_{xx}+2ip\vert p\vert^2\nonumber\\
&&\qquad+\left[\frac{i}{2}(1-\alpha')+pT(t)-\alpha\right]q\vert q\vert^2+ipT(t)+iqA(t)\nonumber\\
&&\qquad-\frac{\alpha}{2}\rho q\int_0^x\left(qq^*_y-q^*q_y\right)dy.\label{RR07}
\eea
The constant coupling parameters in Eq. \ref{NN04} make it possible to cast it as the NHD of NLS
systems with both constant and local coupling parameters, unlike the inhomogeneous HSC case. As a
pointer, the above NHD effectively amounts to identifying $T(t)=-A(t)$ as the drag itself, symbolizing
non-integrability.

\paragraph*{A particular duality:}Considering the most general NHD at ${\cal O}\left(\lambda^0\right)$ 
with local deformation parameters $f_{3,\pm}$, from Eq.s \ref{R01} and \ref{RR06} it is evident that
the system goes through a {\it local} scaling $q\to p=\rho q$ with $\rho=\rho(x,t)$ in general. Thus
`localization' of the coupling parameter puts it on the same footing as the dynamical variable,
{\it i. e.}, $\rho(x,t)$ and $q(x,t)$ could trade places. This essentially is a general property of
localization of NLS parameters starting with the Lax pair construction (Eq.s \ref{13}). Therefore an
NLS system having a local coupling, which can experimentally be realizable,
always corresponds to a dual NLS system made of the coupling parameter itself, with $q(x,t)$ assuming
the role of corresponding coupling. This additionally implies that both continuum cases of inhomogeneous HSC
and vortex filament with drag can be thought as products of deformation of any one of these sectors,
so as any other NHD of the {\it localized} NLS system. This aspect will be studied elsewhere.

\section{Properties of NHD of NLS system}
The NHD of NLS system displays generic properties owing to both spectral algebra of the
integrability structure and generic localization of the coupling parameter. Here we discuss two important
ones among them. Noticeably, a local NLS coupling can only be achieved by considering a generic
${\cal O}\left(\lambda^0\right)$ NHD, or by extending the Lax pair itself to include that
particular NHD. As a result, from the perspective of all possible NHDs, a generic analytical structure
emerges with coupling-localization signifying the semi-classical aspect.

\subsection{The spectral bound}
As mentioned previously the spectral order of perturbation $\delta B$ crucially effects the desired
localization of coupling parameters $(\rho,\eta)$. We now
demonstrate that the NHDs leading to Eq. \ref{N7} are restricted to a very specific
range of spectral parameter powers which is in one-to-one correspondence with the
NHD of HSC itself. The example of XXX model is considered for this demonstration as the integrability
structure of this parent system is also well-understood. \\

In the continuum sector (subsection \ref{2.1}) only ${\cal O}\left(\lambda^{0,-1}\right)$
deformations can lead to the desired result. Among them the ${\cal O}\left(\lambda^0\right)$ deformation
keeps the parameters ($\rho,\eta$) local at ${\cal O}\left(\lambda^1\right)$ sector of ZCC, whereas the
${\cal O}\left(\lambda^{-1}\right)$ deformation allows {\it free} local variables at
${\cal O}\left(\lambda^{-1}\right)$ sector of the same. The desired deformed EOM is obtained in the
${\cal O}\left(\lambda^0\right)$ sector of the ZCC. Since the ZCC for pure NLS system itself is
limited within ${\cal O}\left(\lambda^{0,1}\right)$ the above s the only combination yielding the
desired result. Any other deformation of ${\cal O}\left(\lambda^{n<-1}\right)$ or
${\cal O}\left(\lambda^{n>0}\right)$ will not contribute.

\paragraph*{}To see that a corresponding spectral restriction on NHD exists also in the discrete
analogue, {\it i. e.} in case of HSC, let us consider the semi-Classical limit of the corresponding
EOM,

\be
S_t=\frac{1}{2i}\left[S,S_{xx}\right].\label{1}
\ee 
Here $S=t_i\sigma_i,~ i=1,2,3$ is the matrix representation of the semi-classical spin vector ${\mathbf t}$
in the $SU(2)$ subspace $\{\sigma_i\}$. This system is solvable, having the corresponding Lax pair
\cite{01,LPE}:

\be
U=i\lambda S,\quad V=2i\lambda^2 S-\lambda S_xS,\quad S^2={\mathbb I},\label{RR2}
\ee
that ensures integrability through the ZCC: $F_{tx}=U_t-V_x+\left[U,V\right]=0$.\\

For a general NHD manifested by deforming the temporal Lax component,

\be
V\rightarrow V_d=V+\delta V,\quad \delta V=\frac{i}{2}\sum_n\lambda^n{\boldsymbol\alpha}^{(n)}\cdot{\boldsymbol\sigma},\quad n\in{\mathbb I},\label{5}
\ee
with coefficients $\alpha_i^{(n)}$ are coefficients, the deformed EOM is obtained as:

\bea
&&S_t=\frac{1}{2i}\left[S,S_{xx}\right]+\frac{1}{2}\Lambda^{(1)}_x-i\left[S,\Lambda^{(0)}\right],\nonumber\\
&&\Lambda^{(n)}:={\boldsymbol\alpha}^{(n)}\cdot{\boldsymbol\sigma}.\label{N02}
\eea
Only the ${\cal O}\left(\lambda^{0,1}\right)$ deformations contribute to the EOM. Modulo the common
factor of $\lambda$ in the Lax pair in Eq. \ref{RR2}, this contribution is confined to
${\cal O}\left(\lambda^{0,-1}\right)$, identical to the continuum analogue (NLS) over the
Hasimoto map. For all $n\neq 0,1$, generic recursive constraints of the form,

\be
\Lambda^{(n)}_s-i\left[S,\Lambda^{(n-1)}\right]=0,\label{7}
\ee
are obtained. Therefore there is a bound in the spectral hierarchy about $n\in(0,1)$
(or $n\in(-1,0)$) that exclusively contributes to the NHD, both in the discrete case and also in the 
continuum limit. This in turn identifies the NHD of Eq.s \ref{R02} and \ref{R04} to be the only
possible one yielding the desired result in Eq. \ref{N7}. From the last section, this is also true
in case of vortex filament motion with drag.

\subsection{The semi-Classical nature}
The implication of generalizing the NLS system to incorporate inhomogeneity by localizing the coupling
parameter is reflected in the unique condition Eq. \ref{R07} absent otherwise. This may be identified
as a semi-Classical signature that an {\it inhomogeneous} discrete system would display in the continuum
limit as a result of quantum coherence among individual spins. As a result phase and amplitude of
the solution $q$ get correlated beyond the EOM.\\

To isolate localization of coupling as the cause of this semi-Classical character,
we consider a system {\it without} NHD. It implicates $\rho=\rho(t)$ from Eq. \ref{N013} as
$f_{3,\pm}=0$. Then the last of Eq.s \ref{N012} identify $\eta(x,t)=-\vert\rho\vert^2\equiv\eta(t)$.
This is the most general implication of the Lax construction (Eq.s \ref{13}) allowing for a
time-dependent coupling. Although no more integrable it would correspond to a
time-dependent ferromagnetic parameter in the HSC\footnote{Such a time-dependent HSC may define
a spin density wave or a spin chain under time-varying magnetic field. The continuum NLS analogue
then would imply gradual change from strong to weak coupling and {\it vice versa}, which is experimentally
realized through Feshbach resonance. This could be the mechanism depicted by the `new' NLS Eq. \ref{R01}
of unit coupling, obtained through {\it restricted} NHD of Eq. \ref{N011}.}. The corresponding dynamics
that finally arises,

\be
p_t-ip_{xx}+2ip\vert p\vert^2=0,\quad p=\rho q,
\ee
is of the defocusing NLS type with unit coupling. In general, such time-dependence may correspond to
some additional dynamics of the Frenet-Serret curve in the similar sense of drag, observed for the
vortex filament in the last section.\\

Essentially in this most general undeformed NLS equation,
the phase and amplitude of the complex solution $q=\vert q\vert^{i\theta}$ are related only
through the EOM. This is confirmed by the re-appearance of NLS dynamics with a scaled variable. A
HSC with time-dependent ferromagnetic parameter still maintains its Classical nature since the
corresponding continuum limit still represents a {\it spatially} extended object. Unless the coupling
is made spatially local (inhomogeneous), semi-Classical signature cannot be expected. This is not the
case for Eq. \ref{N011} leading to the scaled defocusing case of Eq. \ref{R01}.\\

The NLS coupling is made local through the {\it additional}
constraint of Eq. \ref{R07} which is a fourth-order differential equation beyond the dynamics of
Eq. \ref{N7}. As the deformed system is integrable \cite{N3} the exact solutions in the space
of functions are now restricted by this constraint. This is the extra phase-amplitude correspondence.
Additional NHDs will further restrict this sector through even higher constraints, forming a
{\it hierarchy} \cite{KunduNHD}. The presently constrained subspace physically incorporates
semi-Classical dynamics. The explicit expression of Eq. \ref{R07} in terms of $\theta$ and
$\vert q\vert$ is straightforward to obtain, but is tediously long to express here.\\

The corresponding condition for vortex filament motion with drag is given by Eq. \ref{RR06},
through the general constraint structure of Eq. \ref{R07}. Similar to continuum inhomogeneous HSC,
the exact equation will look tediously long. This system embodies additional local dynamics in the
Frenet-Serret representation itself \cite{NN1,NN2}.\\

Therefore the introduction of spatial locality to the NLS coupling constant invokes
semi-Classical behavior, manifesting as additional phase-amplitude coupling. The particular form
of the same is determined by the non-holonomic constraints that impose the desired
integrodifferential generalization of the NLS system.


\section{Conclusion and Further Possibilities}
The Hasimoto analogues of continuum inhomogeneous XXX HSC and quantum vortex filament with drag are
shown to be two particular non-holonomic NLS systems. This identification is exclusive to a particular spectral
range $\lambda^{0,-1}$ representing a spectral bound. Such deformations can also be interpreted
in the corresponding Frenet-Serret manifold, yielding
additional amplitude-phase correspondence with semi-Classical nature that owes to
localization effects. The usual NLS system is inherently `Classical' being a mean-field
description of homogeneous XXX model and vortex filament without drag. Their precise NHDs
are further characterized by the integrability of the inhomogeneous HSC (semiholonomic)
and non-conservativeness of the filament-drag case. The exact form of these deformations
are strictly subjected to the fourth-order constraint equations of Eq. \ref{R07}. We leave the
investigation of such exact solutions to the recent future.

\paragraph*{}It is natural to ask if more general spin systems could correspond to non-holonomic
differential equations. The immediate candidates for this are XXZ/XYZ spin chains, with additional
{\it anisotropy} in the $SU(2)$ subspace:

\be
H=-J\sum_{i=1}^N\sum_{a=1}^3\left[\zeta^a{\cal S}^a_i\cdot{\cal S}^a_{i+1}\right].\label{N019}
\ee
The XXZ system corresponds to $\zeta^1=\zeta^2\neq\zeta^3$, whereas $\zeta^1\neq\zeta^2\neq\zeta^3$
results in the XYZ model. One can subsequently construct:

\bea
&&S=\left( \begin{array}{cc} \zeta^3t_3 & \zeta^1t_1-i\zeta^2t_2 \\ \zeta^1t_1+i\zeta^2t_2 & -\zeta^3t_3 \end{array} \right)\equiv \sum_{i=1}^3T_i\sigma_i,\nonumber\\
&&T_i=\zeta^it_i,\label{N20}
\eea
with ${\boldsymbol T}$ being the Frenet-Serret tangent having an {\it additional} constraint in $\zeta^i$s.
Then the above procedure of NHD will go through owing to the decoupled structure
${\cal M}={\mathbb R}\otimes SU(2)$ of the complete vector space. However as a down-side this
persistent anisotropy prohibits the usual Hasimoto map. Possibly more complicated
Frenet-Serret curves may represent the semi-Classical limits of such systems. In a wider sense a
general class of discrete-to-continuous correspondence can possibly be obtained, including vortex
filament motion with drag, resulting in a different class of NHDs through the Hasimoto map. We
leave such attempts for the future.

\begin{figure}[h]
\centering
\resizebox{1\columnwidth}{!}{%
  \includegraphics{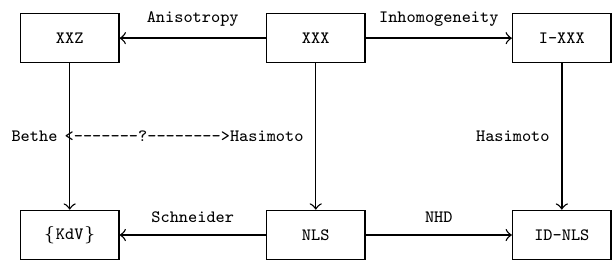}
}
\caption{Schematics of different HSCs corresponding to continuum integrable systems. Here I-XXX stands for inhomogeneous
XXX and ID-NLS for integrodifferential NLS systems. The right half represents the present results, whereas the left half
summarizes the possibilities regarding {\it anisotropic} generalizations. \{KdV\} includes quantum (modified) KdV systems.
The possible connection between Hasimoto map and Bethe ansatz is represented by the dashed line.}
\label{FC}
\end{figure}

\paragraph*{}An indirect approach for anisotropic HSC systems could exploit the weak correspondence
between NLS and KdV systems through the Schneider map \cite{4,5}. A generalization in the KdV side
to incorporate the inhomogeneity in Eq. \ref{N7} may serve the purpose as these two systems also
mutually complement over quasi-integrable deformations \cite{Own}. Further, as the Bethe Ansatz for
the quantum (modified) KdV equation \cite{6,7,8} is a continuum limit of the XXZ model \cite{5}, the 
proposed generalized Hasimoto map may lead not to NLS, but to the KdV system. Then the particular
Bethe Ansatz could be related to the Hasimoto map. A schematic representation of this scheme is given
in Fig. \ref{FC}.

\vskip 0.6cm
\noindent{\it Acknowledgement:} The authors are grateful to Professors Luiz. A. Ferreira and Wojtek
J. Zakrzewski for their encouragement and useful discussions. The research of KA is supported by
the T\"UBITAK 2216 grant number 21514107-115.02-124285 of the Turkish government.
The research of PG was partially supported by FAPESP through
Instituto de Fisica de S\~ao Carlos, Universidade de Sao Paulo with grant number 2016/06560-6.

\end{document}